\documentclass[pre, 12 pt]{revtex4}
\usepackage[cp1251]{inputenc}
\usepackage[english,russian]{babel}

\usepackage{graphicx}
\usepackage{dcolumn}
\usepackage{eucal}
\usepackage[dvips]{epsfig}
\usepackage{amssymb}
\usepackage{amsmath}

\begin{document}


\newcommand{\bs}{\boldsymbol}
\newcommand{\mbb}{\mathbb}
\newcommand{\mcal}{\mathcal}
\newcommand{\mfr}{\mathfrak}
\newcommand{\mrm}{\mathrm}

\newcommand{\ovl}{\overline}
\newcommand{\p}{\partial}

\renewcommand{\d}{\mrm{d}}
\newcommand{\lap}{\triangle}

\newcommand{\lan}{\bigl\langle}
\newcommand{\ran}{\bigl\rangle}

\newcommand{\bse}{\begin{subequations}}
\newcommand{\ese}{\end{subequations}}

\newcommand{\be}{\begin{eqnarray}}
\newcommand{\ee}{\end{eqnarray}}

\newcommand{\ga}{\alpha}
\newcommand{\gb}{\beta}
\newcommand{\gc}{\gamma}
\newcommand{\gd}{\delta}
\newcommand{\gr}{\rho}
\newcommand{\eps}{\epsilon}
\newcommand{\veps}{\varepsilon}
\newcommand{\gs}{\sigma}
\newcommand{\gf}{\varphi}
\newcommand{\go}{\omega}
\newcommand{\gl}{\lambda}

\renewcommand{\l}{\left}
\renewcommand{\r}{\right}

\author{S.A. Trigger$^{1),2)}$}
\title{Equilibrium radiation in a plasma medium with spatial and frequency dispersion}

\address{
$^{1)}$ Joint Institute for High Temperatures, Russian Academy of Sciences, Moscow 125412, Izhorskaya str. 13, build. 2, Russia\\
$^2)$ Institut f\"ur Physik, Humboldt-Universit\"at zu Berlin,
Newtonstra{\ss}e 15, D-12489 Berlin, Germany\\
е-mail: satron@mail.ru}

\begin{abstract}

Examination of equilibrium radiation in plasma media shows that the spectral
energy distribution of such radiation is different from the Planck equilibrium radiation. Using the approach of quantum electrodynamics the general relation
for the spectral energy density of equilibrium radiation in a system of charged particles is found. The obtained result takes into account the influence of plasma on equilibrium radiation through the explicit transverse dielectric permittivity which takes into account spatial and frequency dispersion, as well as the finite collisional damping. For the limiting case of an infinitesimal damping the result coincides with the known expression.

\end{abstract}

\maketitle

The spectral energy distribution of the equilibrium radiation (SEDER) has been established by M. Planck [1] and corresponds to an idealized model of an absolutely black body, which exists in a cavity filled with radiation and bounded by an absolutely absorbing substance. It is assumed that the radiation is in thermodynamic equilibrium with the substance, although the effects of the interaction of photons with the substance bounding the cavity are not considered [2].
The Planck distribution
\begin{eqnarray}
e_P(\omega)\equiv\frac {d E(\omega)}{d\omega} = \frac{V\hbar}{\pi^2 c^3}\frac{\omega^3}{\exp(\hbar\omega/T) -1}, \label{F1}
\end{eqnarray}
is usually associated with the consideration of a macroscopic body in thermal equilibrium with the surrounding "black"\, radiation (see [3,4] for more detail and the references therein). In (1) $ V $ is the volume in which the radiation is enclosed, $ T $ is the temperature of the medium (in energy units) surrounding this volume, $ c $ is the speed of light in vacuum.
It is worth noting, that much attention was paid to the study of the optical properties of various plasma systems [5–8], but not to the SEDER investigation.

The SEDER in the substance itself, which is in a state of thermodynamic equilibrium with radiation, is considered much less frequently (see [9–18] and the literature therein).
In the presence of a substance, the corresponding result differs from the Planck distribution, which corresponds to an ideal photon gas [2]. In addition, the presence of at least a small amount of matter is necessary for the possibility of obtaining equilibrium radiation, since the direct interaction between photons is negligible [2].

The main efforts for last two decades has been devoted to the problem of accounting of the spatial dispersion of the properties of plasma medium (in parallel with the frequency dispersion) on the modification of SEDER [9],[15],[16]. The solution of this problem is essential, in particular, for astrophysical applications [12], [17], where the role of spatial dispersion on the dielectric permittivity is especially essential for the relativistic case [19].  However, the general solution of this problem is still absent and the various approaches lead to the different results. The semi-macroscopic approach [9] in the limit of negligible spatial dispersion of the dielectric permittivity cannot reproduce the classical Brillouin result for plasma medium (see, e.g., [2], [12]). The same problem appears in the rigorous consideration on the basis of quantum electrodynamics (QED) approach [17], [20], where the peculiarities of the SEDER for the asymptotically high and low frequencies have been found. The more elaborated semi-macroscopic result is developed in [16], where the Brillouin limit is fulfilled. However, the final result in this case cannot be expressed through the transverse dielectric permittivity only.

As shown in this letter, to calculate the SEDER correctly the rigorous QED approach in plasma medium should be extended by accounting the additional term in the Hamiltonian, which was previously skipped in  [15], [20]. For simplicity we consider below the case of non-relativistic particles. The obtained result recovers the Brillouin limit in the case of negligibility of spatial dispersion and at the same time permits to find the closed relation for SEDER in presence of the spatial and frequency dispersion of the transverse dielectric permittivity. The final expression for the SEDER contains only the transverse dielectric permittivity of disordered plasma medium with arbitrary interaction between charged particles.

As shown in [18] (on the basis  of [15], [20]) the part of the SEDER in the presence of plasma can be written in the form
\begin{eqnarray}
e^{(1)}(\omega)=\frac{V\hbar \omega^3}{\pi^2 c^3}\coth\left(\frac{\hbar\omega}{2T}\right)\frac{c^5}{\pi \omega}\int_0^\infty d k k^4 \frac{{\textrm {Im}}\varepsilon^{tr}(k,\omega)}{(\omega^2 {\textrm {Re}}\varepsilon^{tr}(k,\omega)-c^2 k^2)^2+\omega^4({\textrm {Im}}\varepsilon^{tr}(k,\omega))^2}
 \label{F2}
\end{eqnarray}
Here $\varepsilon^{tr}(k,\omega)$ is the transverse DP for the non-relativistic plasma system. This part of the SEDER corresponds to the averaged term in the Hamiltonian of interacted radiation and plasma
\begin{eqnarray}
\tilde H_f= \sum_{k,\lambda}\omega_k c^+_{\textbf{k},\lambda} c_{\textbf{k},\lambda}; \qquad  \omega_k=c \mid\textbf{k}\mid; \qquad [c_{\textbf{k},\lambda}c^+_{\textbf{k}',\lambda'}]_{-}=\delta_{\textbf{k},\textbf{k}'}
\delta_{\lambda,\lambda'},
\label{F3}
\end{eqnarray}
where $c_{\textbf{k},\lambda}$ and $c^+_{\textbf{k}',\lambda'}$ are the creation and annihilation operators for photons with momentum $\textbf{k}$ and polarization $\lambda=1,2$. Eq. (3) takes into account the influence of plasma particles on the photon Green function [21].
Evidently, expression (2) is essentially positive due to inclusion of the zero vacuum fluctuations.

In the limit of zero collisional and Landau damping, when ${\textrm {Re}}\varepsilon^{tr}(k,\omega)\rightarrow {\textrm {Re}}\varepsilon^{tr}_0(k=0,\omega)= 1-\sum_a (\omega_{p a}^2/\omega^2)\equiv \varepsilon_0(\omega)$ (${\textrm {Im}}\varepsilon^{tr}(k,\omega)\rightarrow +0$;\;index $a$  indicates the particle species, $\omega_{p a}$ is the plasma frequency for particles of species $a$) Eq. (2) leads to the expression for the SEDER in the transparent plasma $e_{trans}^{(1)}(\omega)$
\begin{eqnarray}
e_{trans}^{(1)}(\omega)=
\frac{V\hbar \omega^3}{2\pi^2 c^3}\left(1-\frac{\omega_p^2}{\omega^2}\right)^{3/2}\theta(\omega-\omega_p)+ \frac{V\hbar \omega^3}{\pi^2 c^3}\frac{1}{\exp(\hbar\omega/T)-1}\left(1-\frac{\omega_p^2}{\omega^2}\right)^{3/2}\theta(\omega-\omega_p),
 \label{F4}
\end{eqnarray}
Here $\omega_p^2=\sum_a \omega^2_{p a}$ is the full plasma frequency.
The first term in (4) corresponds to the modified in plasma zero oscillations [18]. The second one is the modified by the presence of the transparent plasma Planck distribution. However, the classical result for SEDER in the non-dissipative and non-damping plasma, going back to Brillouin, is proportional to square root of $\varepsilon_0(k=0, \omega)$. In contradiction with this result the respective term in (4) is proportional to $\varepsilon^{3/2}_0(k=0, \omega)$.

Let us consider now the term
\begin{eqnarray}
\hat {V}= -\frac{1}{c}\int d^3x \textbf{a}(x)\textbf{j}(x)
\label{F5}
\end{eqnarray}
in a full Hamiltonian of photons and charged particles. Taking into account the part of the operator of  inner current density $\textbf{j}(x)$ proportional to the operator $ \textbf{a}(x)$
\begin{eqnarray}
\textbf{j}(x)= \sum_a \frac{e^2_a}{m_a c} \textbf{a}(x)\psi^+_a(x)\psi_a(x)
\label{F6}
\end{eqnarray}
we arrive to the averaged energy related to Eq. (5)
\begin{eqnarray}
\langle V\rangle=
\frac{1}{c}\int d^3x \sum_a \frac{e^2_a}{m_a c} \langle\textbf{a}^2(x)\rangle \langle\psi^+_a(x)\psi_a(x)\rangle
  \label{F7}
\end{eqnarray}
In (6) and (7) the values $\psi^+_a(x)$ and $\psi_a(x)$ are the operators
of creation and annihilation of particles of species $a$. For homogeneous system the
simple calculation of the average $\langle\textbf{a}^2(x)\rangle$ leads to the result for the additional term in the SEDER $e^{(2)}(\omega)\equiv\langle V\rangle$
\begin{eqnarray}
e^{(2)}(\omega)=
\sum_a \frac{4 \pi e^2_a n_a}{m_a}  \sum_{k} \frac{\hbar}{\omega_k}\{2 \langle c^+_{\textbf{k}}c_{\textbf{k}}\rangle+1\}=V \int \frac{ d^3 k}{(2\pi)^3}\sum_a \frac {\hbar\omega^2_{pa}}{\omega_k}[2 f(k)+1] \label{F8}
\end{eqnarray}
where $n_a$, $f(k)$ are the density of particles of species $a$ and the photon distribution function in plasma medium respectively.

Using the relation between $f(k)$ and the photon retarded Green function in Coulomb gauge $D^R (k,\omega)$ [15, 22] for the homogeneous Coulomb system
\begin{eqnarray}
f(k)=- \left[\frac{1}{2}+\frac{k}{2\pi c}\int_0^\infty  \frac{d\omega}{2\pi}\coth \left(\frac{\hbar\omega}{2T}\right)\textrm{Im} D^R(k,\omega)\right],\label{F9}
\end{eqnarray}
where
\begin{eqnarray}
D^R(k,\omega)=\frac{4\pi c^2}{\varepsilon^{tr}(k,\omega)\omega^2-k^2c^2}.\qquad \textrm{Im} D^R(k,\omega)=-\frac{4\pi c^2 \omega^2 \textrm{Im} \varepsilon^{tr}(k,\omega)}{\mid\varepsilon^{tr}(k,\omega)\omega^2-k^2c^2\mid^2}\label{F10}
\end{eqnarray}
we arrive at the expression for $e^{(2)}(\omega)$ (with $\omega_k=c k$)
\begin{eqnarray}
e^{(2)}(\omega)=2V \int \frac{ d^3 k}{(2\pi)^3}\sum_a \frac {\omega^2_{pa}}{\omega_k}[f(k)+\frac{1}{2}]=\nonumber\\- 2V \int \frac{ d^3 k}{(2\pi)^3}\sum_a \frac {\omega^2_{pa}}{2\pi c^2}\int_0^\infty  \frac{d\omega}{2\pi}\coth \left(\frac{\hbar\omega}{2T}\right)\textrm{Im} D^R(k,\omega)\;\;
\label{F11}
\end{eqnarray}
Taking into account Eq. (11) and the definition of the relation between the energy of electromagnetic field $E$ in plasma and the full SEDER $e(\omega)=e^{(1)}(\omega)+e^{(2)}(\omega)$
\begin{eqnarray}
E \equiv \int_0^\infty d\omega e(\omega)
\label{F12}
\end{eqnarray}
we find finally for $e^{(2)}(\omega)$
\begin{eqnarray}
e^{(2)}(\omega)=V
\frac{\hbar\omega^2 \omega^2_{p}\coth \left(\frac{\hbar\omega}{2T}\right)}{\pi^3} \int_0^\infty d k k^2\frac{\textrm{Im} \varepsilon^{tr}(k,\omega)}{\mid\omega^2\varepsilon^{tr}(k,\omega)-k^2c^2\mid^2}\;\;
\label{F13}
\end{eqnarray}

After summation of the terms (2) and (13) we arrive to the full SEDER which takes into account the spatial and frequency dispersion of the plasma medium
\begin{eqnarray}
e(\omega)=V\frac{\hbar \omega^2 \omega_p^2}{\pi^3}\coth\left(\frac{\hbar\omega}{2T}\right)\times\nonumber\\
\int_0^\infty d k k^2\left(1+ \frac{c^2k^2}{\omega_p^2}\right) \frac{{\textrm {Im}}\varepsilon^{tr}(k,\omega)}{(\omega^2 {\textrm {Re}}\varepsilon^{tr}(k,\omega)-c^2 k^2)^2+
\omega^4({\textrm {Im}}\varepsilon^{tr}(k,\omega))^2}
 \label{F14}
\end{eqnarray}

As an example let us consider the case of transparent medium under condition of the absence of the spatial dispersion of DP. In this case
\begin{eqnarray}
 \left(\frac{{\rm Im} \varepsilon^{tr}(k,\omega)}{({\rm Re}(k,\omega) \varepsilon^{tr}\omega^2-c^2k^2/\omega^2)^2 + ({\rm Im} \varepsilon^{tr}(k,\omega)^2}\right)\mid_{k \rightarrow 0; {\rm Im} \varepsilon^{tr}(k,\omega)\rightarrow+0}\rightarrow \pi\delta(1-c^2k^2/\omega^2), \label{F15}
\end{eqnarray}
where $\delta(x)$ is the Dirac delta function. After the simple calculation we find the answer
\begin{eqnarray}
e_{trans}(\omega)=
\frac{V\hbar \omega^3}{2\pi^2 c^3}\left(1-\frac{\omega_p^2}{\omega^2}\right)^{1/2}\theta(\omega-\omega_p)+ \frac{V\hbar \omega^3}{\pi^2 c^3}\frac{1}{\exp(\hbar\omega/T)-1}\left(1-\frac{\omega_p^2}{\omega^2}\right)^{1/2}\theta(\omega-\omega_p),\;\;
 \label{F16}
\end{eqnarray}
with the square root of DP $\varepsilon_0(\omega)$ instead Eq. (4) for $e_{trans}^{(1)}(\omega)) $. The second term in (18) describes the result equivalent to the classical approach. The first term in (18) defines the modified zero fluctuations [18].

Therefore, the explicit spectral energy distribution of the equilibrium radiation in plasma can be described by use Eq. (14) for the non-transparent dissipative medium. The necessary condition for the further calculations is the use of good approximations for transverse dielectric permittivity of plasma medium.

The problem of separation of the modified zero fluctuations in plasma from general expression (14), as well as the relativistic generalization of Eq. (14) will be considered in a separate work.

The author is thankful to A.M. Ignatov and A.G. Zagorodny for many discussions of the considered problem.

\end{document}